\documentclass[aps,twocolumn,showpacs,prb,superscriptaddress,unsortedaddress]{revtex4}
\usepackage{epsf}
\usepackage{graphicx}
\usepackage{bm}
\usepackage{amsmath}
\usepackage{xcolor}
\usepackage{siunitx}

\newcommand{\BSCO}{Bi$_2$Sr$_2$CaCu$_2$O$_{8+\delta}\,$}
\newcommand{\Tc}{T$_{\text{c}}$~}
\newcommand{\Tcc}{T$_{\text{c}}$}

\begin{document}

\title{Particle-hole symmetry in the pseudogap phase of moderately underdoped cuprate  high temperature superconductors evidenced from joint density of states analysis}
\author{Niraj Kumar Shah}
\affiliation{Department of Physics, University of Virginia, Charlottesville, VA 22904}         
\author{Junjing Zhao}
\affiliation{Department of Physics, University of Virginia, Charlottesville, VA 22904}
\author{Utpal Chatterjee}
\thanks{uc5j@virginia.edu (U.C.)}
\affiliation{Department of Physics, University of Virginia, Charlottesville, VA 22904}

\date{\today}

\begin{abstract}
In conventional superconductors, the energy scale associated with the superfluid stiffness is much larger compared to the pairing energy and hence, the superconducting transition temperature (\Tcc) is entirely dictated by the superconducting (SC) energy gap. The phase rigidity of the SC condensate in unconventional superconductors, on the other hand, can be low enough to enable destruction of superconductivity via phase incoherence and persistence of an energy gap even at the absence of macroscopic superconductivity above \Tcc. This is considered  a possible mechanism of the pseudogap (PG) state of cuprate high temperature superconductors (HTSCs). We have investigated the electronic energy ($\omega$) and momentum-separation vector ($\mathbf{q}$) dependence of the joint density of states (JDOS), derived from the autocorrelated Angle Resolved Photoemission Spectroscopy (ARPES) data, from moderately underdoped \BSCO HTSC samples at temperatures below and above \Tcc. We found that $\mathbf{q}$-space structure of the constant $\omega$ JDOS intensity maps and the dispersions of the JDOS peaks are essentially the same both below and above \Tcc. Furthermore, the dispersions of the JDOS peaks above \Tc  are particle-hole symmetric. These observations evince similarity between the nature of the energy gap below and above $\text{T}_{\text{c}}$, which supports preformed pairing scenario for the PG state at least in the moderately underdoped regime.
\end{abstract}

\maketitle
\section*{Introduction} \vspace{-0.1in}
Studies over the past four decades have established the key common features in the phase diagrams of cuprate HTSCs\----such as, Mott insulating state with a long range antiferromagnetic order in parent compounds, quantum melting of the Mott insulating state via carrier doping ($\delta$) leading to a dome-like \Tc vs $\delta$ curve, and the enigmatic PG state with an energy gap persisting up to a temperature $\text{T}^{*} (\delta)$ that can be substantially higher than  $\text{T}_{\text{c}}(\delta)$  over a range of values of \text{$\delta$} \cite{R_HTSC_1, R_HTSC_2, R_HTSC_3, R_HTSC_4, R_HTSC_5, R_HTSC_6, R_HTSC_7, R_HTSC_8, R_HTSC_9, R_HTSC_10}. The consensus in the field is that settling the microscopic origin of the energy gap in the PG state and its relation to the SC gap holds the key to unveil the elusive SC pairing mechanism of the HTSCs. 

Theoretical pictures of the PG state typically come under two broad categories: (i) the PG state is a phase disordered superconductor with an energy gap that is a remnant of the  SC pairing gap \cite{R_PPT_1, R_PPT_2, R_PPT_3, R_PPT_4, R_PPT_5}, and (ii) the PG state stems from some electronic instability \cite{R_DDW, R_OC, R_SDW1, R_SDW2, R_SDW3, R_CDW1, R_CDW2, R_CDW3, R_CDW4}\----charge density wave (CDW), spin density wave (SDW), $d$-density wave (DDW), etc.\----competing with  superconductivity. Even though these two schools of thought are qualitatively different, it is rather striking that both have been invoked to elucidate experimental data from HTSCs. For instance, while terahertz spectroscopy data \cite{R_THZ}, together with the giant Nernst signal \cite{R_NS1, R_NS2} and the anomalous diamagnetism \cite{R_DM} above \Tc have been interpreted as indicators for the presence of preformed pairs in the PG state, data from various scattering probes reveal ubiquitous density wave instabilities\----charge \cite{R_CDW1, R_CDW2, R_CDW3, R_CDW4}, orbital \cite{R_OC} and spin-density wave orders \cite{R_SDW1, R_SDW2, R_SDW3} in the cuprate HTSCs. It is worth noting that the dispersion relationship of a single-particle excitation near the chemical potential ($\mu$) of the system, in which the information on coherence factors is encoded, can provide critical insights to distinguish between these two very different proposals of the PG state. For instance, the momentum eigenstates, known as the Bogoliubov quasiparticles, of a time reversal invariant superconductor, are coherent superpositions of the electron and hole states instead of being just the bare electron states. When this electron-hole mixing occurs with the zero center-of-mass momentum, the minimum energy gap locus in the SC state coincides with the underlying Fermi surface (FS), which is commonly dubbed as particle-hole symmetry (PHS). Therefore, in the precursor pairing scenario of the PG state, such PHS should be visible along the entire FS, while this is not the case for density wave origin of the PG phase.

ARPES data $\text{I}(\text{k}_\text{x},\text{k}_\text{y},\omega)$ represents the momentum-resolved single-particle-density of states (SDOS) at a specific value of $\omega$, where $\text{k}_\text{x}$ and $\text{k}_\text{y}$ are $\text{x}$- and $\text{y}$-components of the in-plane momentum vector ($\mathbf{k}$), respectively. Typically, $\omega$ is referenced with respect to $\mu$ and we will follow this convention throughout this article. A straightforward test of the PHS in a system using ARPES would involve simultaneous tracking of the dispersions and the spectral weights of both the upper (above $\mu$) and lower (below $\mu$) branches of the Bogololibov quasiparticles from $\text{I}(\text{k}_\text{x},\text{k}_\text{y},\omega)$. To this end, there have been numerous ARPES measurements both in the SC and PG states of the cuprate HTSCs \cite{R_JC_BB, R_MATSUL_BB, R_ZX_BALATSKI_PRB, R_MING_BB, R_AK_BB, R_AK_NODE_ARC_PRL, R_UC_NP, R_UC_PNAS, R_JOHNSON_POCKET, R_FELIX_POCKET, R_ZX_HASHIMOTO, R_ZHOU_BISCO2212, R_AK_KONDO, R_KONDO_NODE, R_ZHOU_BISCO_POCKET}. Even though there is an unanimity on the presence of PHS in the SC state \cite{R_JC_BB, R_MATSUL_BB, R_ZX_BALATSKI_PRB}, it remains a much-debated topic in the PG state \cite{R_ZX_BALATSKI_PRB, R_MING_BB, R_AK_BB, R_JOHNSON_POCKET, R_KONDO_NODE, R_ZHOU_BISCO_POCKET, R_ZX_HASHIMOTO, R_ZHOU_BISCO2212}. In this context, focusing on $\omega$- and $\mathbf{q}$-resolved JDOS intensity maps can provide a new perspective to the investigation of PHS. The elastic autocorrelation function $\text{C}(\mathbf{q}, \omega)$ of ARPES intensity  $\text{I}(\text{k}_\text{x},\text{k}_\text{y},\omega)$ at a fixed value of $\omega$ can be written as follows: $\text{C}(\mathbf{q},\omega)=\sum_{\text{k}_\text{x}, \text{k}_\text{y}}  \text{I}(\text{k}_\text{x},\text{k}_\text{y},\omega) \text{I}(\text{k}_\text{x}+\text{q}_\text{x},\text{k}_\text{y}+\text{q}_\text{y},\omega)$, where the momentum sum is over the first Brillouin zone and $\mathbf{q} \equiv (\text{q}_\text{x}, \text{q}_\text{y})$  \cite{R_OCTET_STM1, R_OCTET_STM2,R_OCTET_STM3,R_OCTET_STM4}. 
$\text{C}(\mathbf{q},\omega)$ at a specific value of $\omega$ can be directly mapped to the $\textbf{q}$-resolved JDOS map at that value of $\omega$, because the peak structures of $\text{C}(\mathbf{q}, \omega)$ directly map out the momentum separation vectors connecting regions with high individual spectral weights of the corresponding SDOS map \cite{R_OCTET_STM1, R_OCTET_STM2,R_OCTET_STM3,R_OCTET_STM4}. Therefore, the $\omega$ vs $|\mathbf{q}|$ relationship of the JDOS peaks can be directly linked to the quasiparticle dispersions. From the above-described mathematical construction of $\text{C}(\mathbf{q},\omega)$, it can be realized that the JDOS analysis involves the momentum sum, which can, in principle, help partially alleviating the data statistics issues if there are any. Consequently, the JDOS analysis can help to enhance and unmask minute details of low-energy electronic excitations that might otherwise be difficult to identify directly from the SDOS data. The goal of this manuscript is to take advantage of this aspect of JDOS analysis to investigate PHS in the PG state. 

The JDOS analysis can also be helpful towards making a direct comparison between the ARPES and Fourier Transformed Scanning Tunneling Spectroscopy (FT-STS) data. Similar to other correlated materials, copper oxide based HTSCs like \BSCO also host quenched disorders in the forms of defects, impurities, etc., which typically cause elastic scattering of the quasiparticles. Such scatterings are considered to be dominant factors for the so-called quasi-particle interference (QPI) patterns, manifesting via spatial modulation in the local density of states. 
Even though the mathematical formulations used for calculating the FT-STS patterns and $\text{C}(\mathbf{q}, \omega)$ are not equivalent, the wave vectors of the QPI patterns from \BSCO  HTSCs have been found to have decent agreement with the $\mathbf{q}$'s  connecting high intensity points of the $\mathbf{k}$-resolved constant $\omega$ SDOS maps, in other words, the peaks of the constant $\omega$ JDOS maps, at least in the optimally doped and moderately underdoped regime \cite{R_OCTET_STM1, R_OCTET_STM2, R_OCTET_STM3, R_OCTET_STM4, R_OCTET_STM5, R_OCTET_STM6, R_OCTET_STM7, R_OCTET_STM8, R_OCTET_STM9}. To resolve this puzzle, there have been extensive theoretical works not only exploring the reasons behind phenomenal success of the  so-called octet phenomenology but also investigating theoretical frameworks beyond  the octet models \cite{R_FT-STS_THEORY1, R_FT-STS_THEORY2, R_FT-STS_THEORY3, R_FT-STS_THEORY4, R_FT-STS_THEORY5, R_FT-STS_THEORY6, R_FT-STS_THEORY7, R_FT-STS_THEORY8, R_FT-STS_THEORY9,R_FT-STS_THEORY10, R_FT-STS_THEORY11}. Nevertheless, the JDOS analysis can potentially play an important role as a bridge between real- and momentum-space spectroscopic techniques.

\begin{figure*}
\includegraphics[width=6.5in]{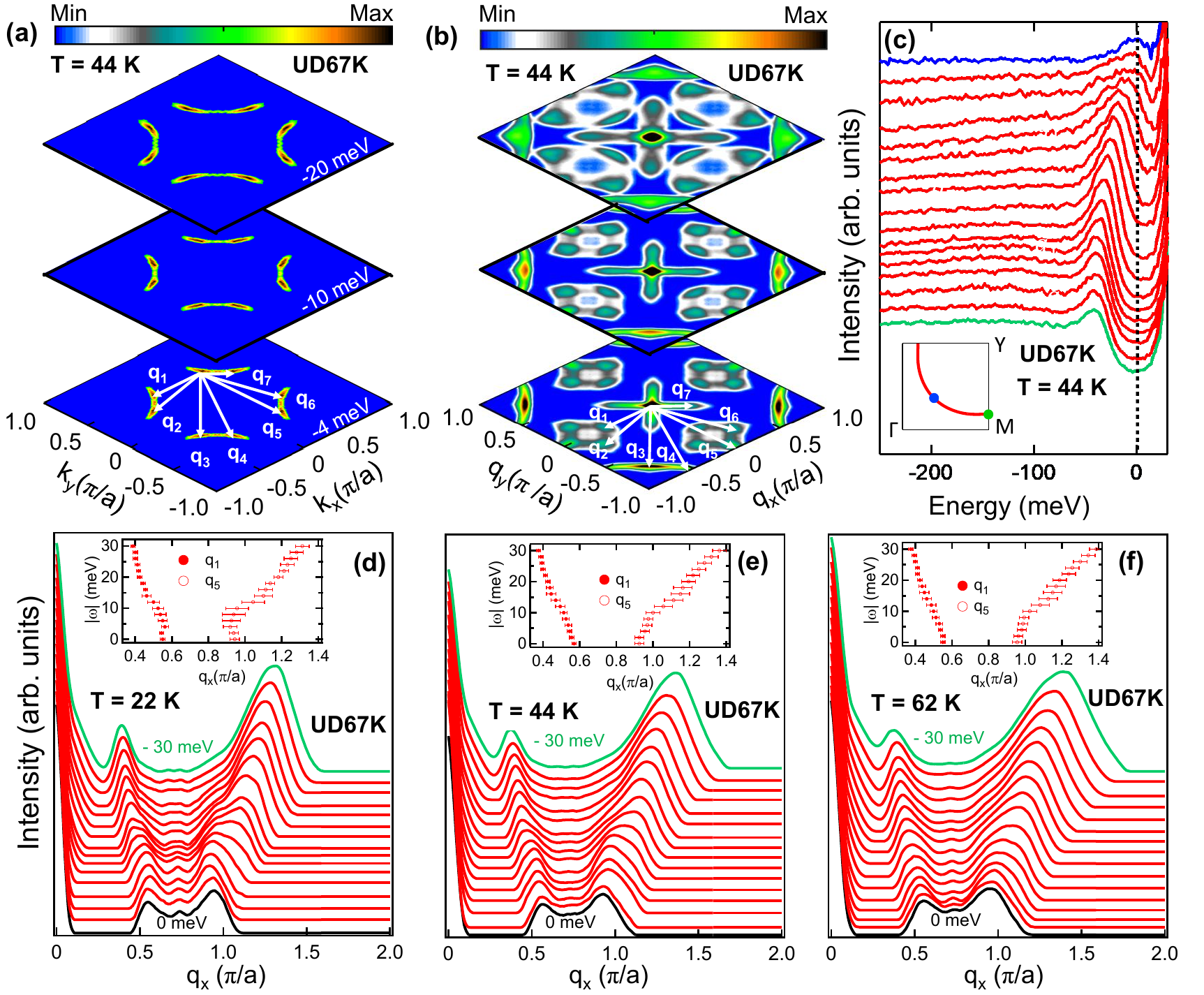}
\caption{\textbf{ Constant $\omega$ SDOS intensity maps, JDOS intensity maps, and dispersions of $|\mathbf{q}_1|$ and $|\mathbf{q}_5|$ in the SC state.} (a) Constant $\omega$ SDOS maps at $\text{T} = 44~\text{K}$ for $\omega =-4$ meV (bottom), $-10$ meV (middle), and $-20$ meV (top). The intensity is anisotropic along the banana-shaped contours of the SDOS maps, with high intensity regions are confined near the ends of the bananas. (b) JDOS maps associated with the SDOS maps in (a). The vectors on the bottom image of (b) indicate local maxima in the JDOS map, which relate to the equivalent vectors separating high intensity points in the bottom image of (a) as predicted by the Octet model. (c) The FF-divided EDCs in the SC state at $\text{T}=44~\text{K}$ along the FS, schematic plot of which is shown in the inset. The green and blue dots on the FS represent the antinode and node, respectively. Strong momentum anisotropy of the SC gap is evident from (c). (d---f) Line cuts of constant $\omega$ JDOS maps along the Cu-O bond direction, i.e., along ($\text{q}_{\text{x}}$,~0) for $-30$ meV$~\le\omega\le~0$ meV, in the SC state at $\text{T}=22~\text{K}$ (d), $44~\text{K}$ (e), and $62~\text{K}$ (f). The black and green curves correspond to $\omega=0$ meV and $\omega=-30$ meV, respectively. Consecutive curves are separated from each other by $2$ meV and the inset of each panel shows $|\omega|$ vs $|\mathbf{q}_1|$ and $|\omega|$ vs $|\mathbf{q}_5|$ plots.}
\label{fig:fig1}
\end{figure*}

\section*{Result}
The summary of our findings is as follows: (i) The energy-momentum structure of the JDOS patterns from moderately underdoped samples do not undergo any significant change as a function of temperature ($\text{T}$) through \Tcc. Just like in the SC state, the peaks of the JDOS intensity maps in the PG state are also dispersive. (ii) The dispersion of the JDOS peaks is particle-hole symmetric in the PG state. 
(iii) The main features of the JDOS patterns and the dispersions of the JDOS peaks in the PG state can be simulated using a simple broadened $d$-wave superconductivity (BDSC) model.

\subsection{JDOS patterns and JDOS peak dispersions in the SC state} \label{Sec2A}

The excitation energy $\text{E}(\mathbf{k})$ of a Bogoliubov quasiparticle in the SC state is given by $\text{E}(\mathbf{k})=\pm \sqrt{\Big(\epsilon(\mathbf{k})\Big)^2+\Big(\Delta(\mathbf{k})\Big)^2}$, where $\epsilon(\mathbf{k})$ is the normal-state band dispersion and $\Delta(\mathbf{k})$ is the momentum-dependent superconducting gap. In cuprate HTSCs, $\Delta(\mathbf{k})$ vanishes at four isolated points in the momentum space, i.e., at the nodes, and consequently, the contour of the SDOS intensity map at $\omega=0$ reduces to four points. As $|\omega|$ is increased, the length of these characteristic banana-shaped contours get extended due to the strong momentum dependence of $\Delta$ and $\epsilon$ \cite{R_OCTET_STM1, R_OCTET_STM2, R_OCTET_STM3, R_OCTET_STM4} . An important feature of these bananas is strongly peaked intensities near their ends because of large value of the curvature of the dispersion, i.e., $\Big| \dfrac{1}{\Delta_{\mathbf{k}}\text{E}(\mathbf{k})} \Big|$. As a result, $\text{C}(\mathbf{q}, \omega)$, which has a one-to-one correspondence to the momentum-resolved JDOS map, is expected to comprise discrete peaks that change as a function of $|\omega|$. To investigate the above-described implications of the $d$-wave SC gap, we focus on data from the UD67K sample below $\text{T}_{\text{c}}$  in Fig. 1. In Fig. 1a, we plot SDOS maps, derived from ARPES data, for $\omega=-4$ meV, $-10$ meV and $-20$ meV at $\text{T}=44~\text{K}$ ($<\text{T}_{\text{c}}=67~\text{K}$). As elaborated above, with increasing $|\omega|$, banana-shaped contours span over larger areas of the Brillouin zone and high intensities remain located near the tips of the bananas. Constant $\omega$ JDOS intensity maps associated with the corresponding SDOS maps in Fig. 1a can be found in Fig. 1b. Expectedly, the constant $\omega$ JDOS maps display well-defined peaks.
As per the Octet model, eight $\mathbf{q}$'s, namely $\mathbf{q}_{i} ~(i = 1,2,..,8)$, are expected to dominate the $\text{C}(\mathbf{q}, \omega)$ intensity maps at each $\omega$ for $|\omega| \leq \Delta_0$, where $\Delta_0$ is the antinodal energy gap. Two of these $\mathbf{q}$'s, namely, $\mathbf{q}_8$ and $\mathbf{q}_4$ are identical due to the symmetry of the system and thus, the peaks in the JDOS map are captured by seven $\mathbf{q}$'s, namely, $\mathbf{q}_{i} ~(i = 1,2,..,7)$. This is precisely what we notice in Fig. 1b, which also confirms that the  locations of the JDOS peaks in $\mathbf{q}$-space evolves dispersively as  $\omega$ is varied from zero to negative values.
 The Fermi function (FF)-divided energy distribution curves (EDCs) along the FS in Fig. 1c verify strong momentum anisotropy of the energy gap\----the $\omega$ location of the peak of the FF-divided EDC at the antinode is furthest from $\mu$ and it gradually moves closer towards $\mu$ as the momentum location of the EDC approaches the node. Expectedly, the peak of the nodal spectra appears at $\mu$. 
 The strongly dispersive nature of the JDOS peaks has been further detailed in Figs. 1d, 1e and 1f.  The dispersion of $|\mathbf q|$ along the Cu–O bond direction (along ($\text{q}_\text{x},0)$), i.e., $|\omega|$  vs $|\mathbf{q}_1|$ and $|\omega|$ vs $|\mathbf{q}_5|$ plots are shown in Figs. 1d, 1e and 1f.  Being consistent with the fact that the SC energy gap increases along the FS from the node to the antinode, $|\mathbf{q}_1|$ gets shorter and $|\mathbf{q}_5|$ gets longer with increasing $|\omega|$ for all measured temperatures below $\text{T}_{\text{c}}$ ($\sim67$ K)\----$22$ K (Fig. 1d), $44$ K (Fig. 1e) and $62$ K (Fig. 1f).

\begin{figure*}
\includegraphics[width=6.5in]{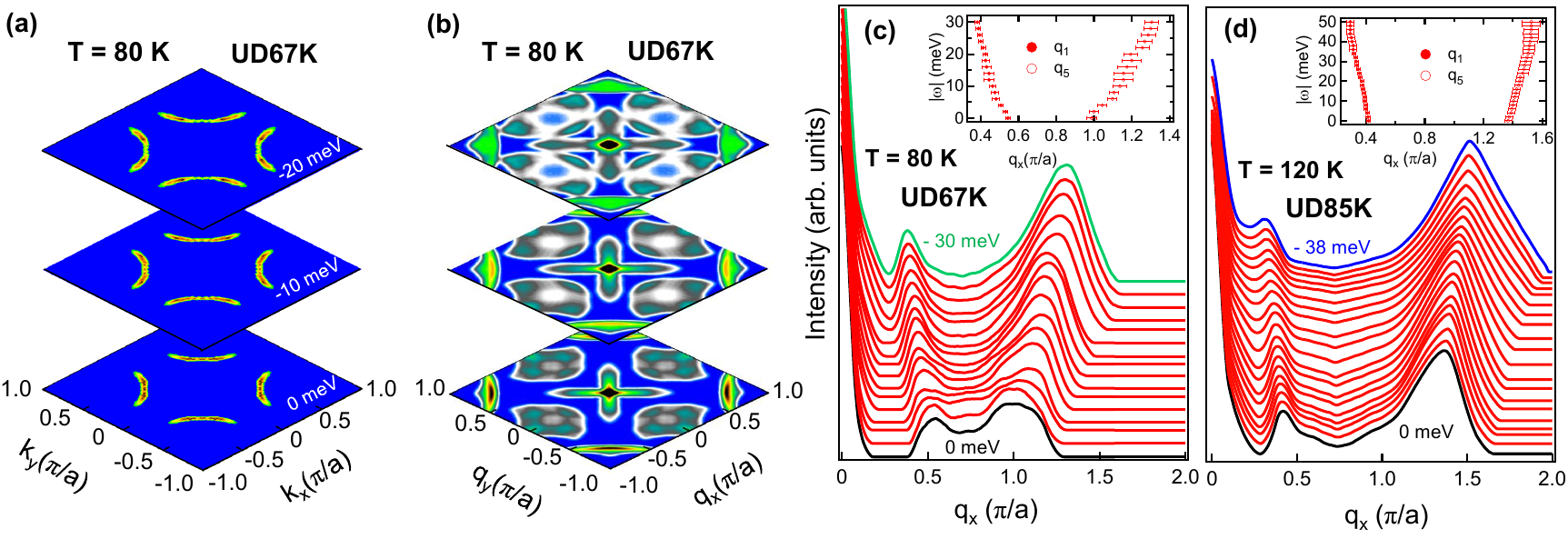}
\caption{\textbf{ Constant $\omega$ SDOS intensity maps, JDOS intensity maps, and dispersions of  $|\mathbf{q}_1|$ and $|\mathbf{q}_5|$ in the PG state.} (a) Constant $\omega$ SDOS maps  in the PG state at $\text{T}=80~\text{K}$ from the UD67K sample for $\omega =0 $ meV (bottom), $-10$ meV (middle), and $-20$ meV (top). (b) JDOS maps associated with the SDOS maps in (a). (c) Line cuts of the constant $\omega$ JDOS patterns from the UD67K sample in the PG state at $\text{T}=80$ K as a function of $\omega$ between $0$ meV and $-30$ meV along the Cu–O bond direction, i.e., along ($\text{q}_{\text{x}}$,~0). The black and green curves in (c) correspond to $\omega=0$ meV and $\omega=-30$ meV, respectively. (d) Same as (c), but from the UD85K sample in the PG state at $\text{T}=120$ K between $0$ meV and $-38$ meV. The black and blue curves in (d) correspond to $\omega=0$ meV and $\omega=-38$ meV, respectively. Consecutive curves are separated from each other by $2$ meV in both (c) and (d). The insets in (c) and (d) show  $|\omega|$ vs $|\mathbf{q}_1|$ and $|\omega|$ vs $|\mathbf{q}_5|$ plots.}
\label{fig:fig2}
\end{figure*}

\subsection{Superconductor-like JDOS patterns and JDOS peak dispersions in the PG state} \label{Sec2B}

A possible signature of the energy gap in the PG state, similar in nature to the SC gap, would be the continued presence of superconductor-like JDOS patterns and JDOS peak dispersions above $\text{T}_{\text{c}}$, which we explore in Fig. 2. We present the constant $\omega$ SDOS maps, derived from ARPES data, for $\omega= 0$ meV, $-10$ meV and $-20$ meV from the UD67K sample at $\text{T}=80~ \text{K}$ ($ > \text{T}_{\text{c}}=67~\text{K}$) in Fig. 2a.  Like in the SC state (Fig. 1a), the contours of the SDOS maps in the PG state also show: (i) the characteristic banana shape, (ii) high spectral intensity in the vicinity of the ends of the bananas, and (iii) an increase in extent with increasing $|\omega|$. The $\mathbf{q}$-space structure of constant $\omega$ JDOS patterns in Fig. 2b are very similar to those in the SC state observed in Fig. 1b. 
Furthermore, we present elaborate views of the dispersions of $|\mathbf{q}|$'s along the Cu–O bond direction, i.e., $|\omega|$ vs $|\mathbf{q}_1|$ and $|\omega|$ vs $|\mathbf{q}_5|$ plots  in Fig. 2c (UD67K sample at $\text{T}=80$ K) and Fig. 2d (UD85K sample at $\text{T}=120$ K). The results are very similar to what have already been found in case  of the SC state in Figs. 1d, 1e and 1f\----$|\mathbf{q}_1|$ gets shorter and $|\mathbf{q}_5|$ gets longer with increasing $|\omega|$. It is worth mentioning that the dispersion of the JDOS peaks in the PG state of the UD67K sample in Fig. 2c is slightly shallower compared to their superconducting counterpart in Figs. 1d, 1e and 1f. Additionally, the dispersion in the PG state of the UD85K sample in Fig. 2d is less pronounced compared to that of the UD67K sample in Fig. 2c. Both observations can be rationalized by considering the impact of the temperature and carrier concentration evolution of the Fermi arcs in the PG state on the JDOS patterns as highlighted in our previous work \cite{R_OCTET_STM3}. Overall, a comparison of data from Figs. 1 and 2 show that the JDOS patterns and the JDOS peak dispersions do not undergo noticeable change as $\text{T}$ changed through $\text{T}_{\text{c}}$, which in turn hint towards energy gap in the PG state being a remnant of the SC gap.
\begin{figure*}
\includegraphics[width=6.5in]{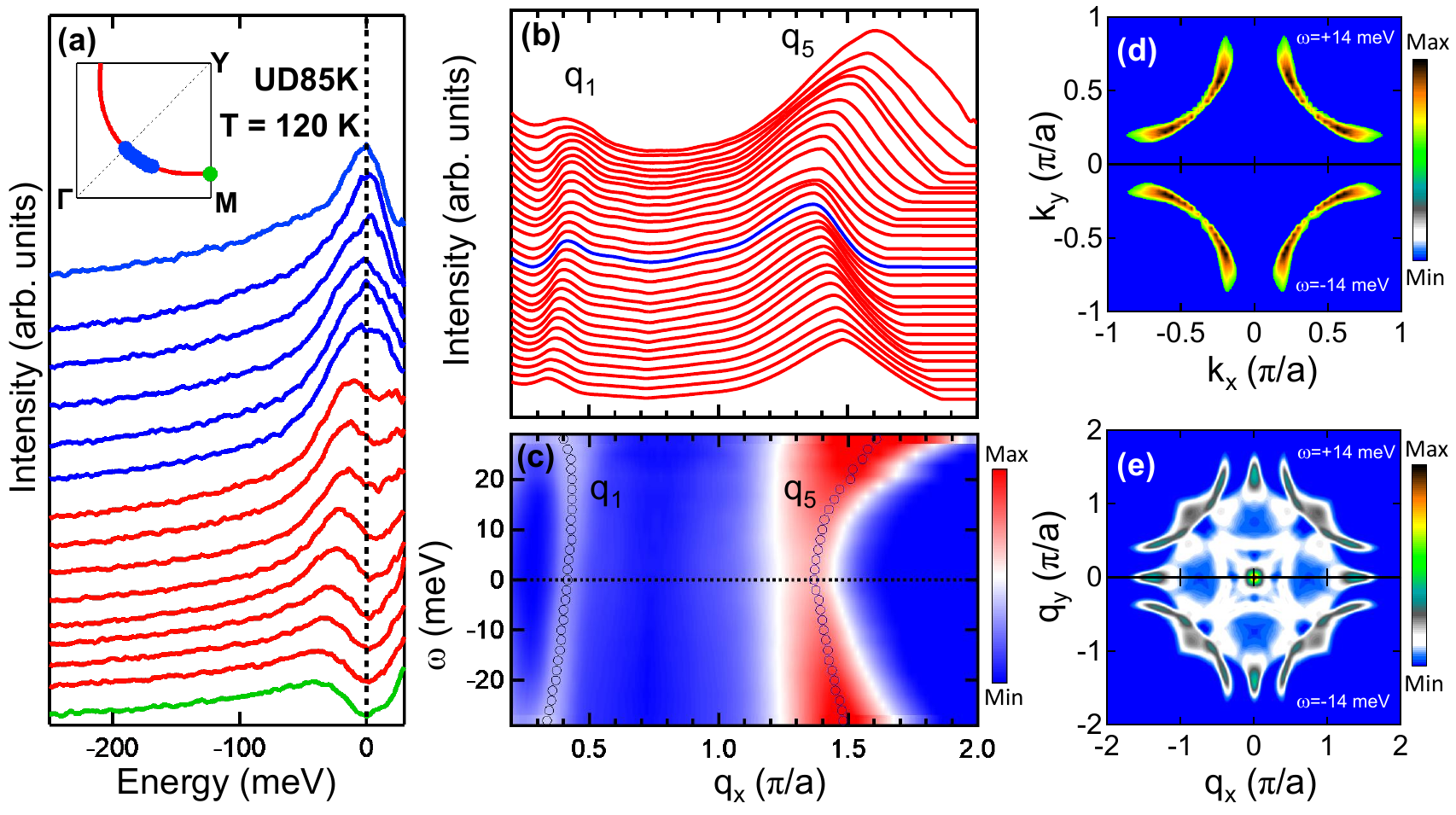}
\caption{\textbf{ Particle-hole symmetry in the PG state.} (a) The FF-divided EDCs from the UD85K sample along the FS at various momenta between the node (topmost curve) and  the antinode (bottommost curve). Multiple single-peaked gapless spectra (blue curves) and gapped spectra (red and green curves) evidence an extended gapless region, i.e., the Fermi arc, around node, while energy gap outside of the Fermi arc. The momenta of the gapless spectra (blue color) are marked by the blue dots on the schematic plot of the FS. The green dot corresponds to  the momentum location of the antinode, while  the green spectrum to the FF-divided EDC at the antinode. The energy location of the minimum of each gapped  EDC is tied to the FS, i.e., it is located at $\mu$. (b) Stacked horizontal cuts of $\text{C}(\mathbf{q}, \omega)$ along the ($\text{q}_\text{x}$, 0) direction for $|\omega|\le 28$ meV. The dispersion curve at each $\omega$ is appropriately normalized for visual clarity. Consecutive curves are separated from each other by $2$ meV. (c) The dispersions of the JDOS peaks of the UD85K sample along the ($\text{q}_\text{x}$, 0) direction as a false-color map in the energy window of $|\omega|\le28$ meV. (b) and (c) indicate  that $|\mathbf{q}_i (\omega)|=|\mathbf{q}_i (-\omega)|$ for $i=1, 5$ within experimental error bars.    
(d) SDOS maps, plotted in one half of the Brillouin zone, for $\omega=-14$ meV (for $|\text{k}_\text{x}|\le(\pi/a)$ and $(-\pi/a)\le\text{k}_\text{y}\le 0$) and $\omega=+14$ meV (for $|\text{k}_\text{x}|\le(\pi/a)$ and $0\le\text{k}_\text{y}\le (\pi/a)$).
 (e) JDOS maps, plotted in part of the $(\text{q}_{\text{x}}, \text{q}_{\text{y}})$ plane, for $\omega=-14$ meV (for $|\text{q}_\text{x}|\le(2\pi/a)$ and $(-2\pi/a)\le\text{q}_\text{y}\le 0$), and $\omega=+14$ meV (for $|\text{q}_\text{x}|\le(2\pi/a)$ and $0\le\text{q}_\text{y}\le(2\pi/a)$).
 JDOS peaks at  $\omega=-14$ meV and  $\omega=+14$ meV seem to mirror each other within experimental error bars, as expected at the presence of PHS.}
\label{fig:fig3}
\end{figure*}
\subsection{The JDOS perspective of PHS in the PG state} \label{Sec2C}

We now turn to the controversial issue of PHS in the PG state, which is critical in settling whether the energy gap in the PG state has pairing origin. A straightforward test for PHS in the PG state can be conducted by examining FF-divided EDCs along the underlying FS above $\text{T}_{\text{c}}$. At the presence of PHS, each FF-divided gapped EDC along the FS is anticipated to exhibit the following: (i) two peaks symmetrically positioned with respect to $\mu$, and (ii) same intensity associated with each peak. The FF-divided EDCs from the UD85K sample at $\text{T}=120\text{K}$ ($>\text{T}_{\text{c}}=85$ K) along the FS are presented in Fig. 3a. As these EDCs are tracked from the node (topmost curve) towards the antinode (bottommost curve), the presence of the Fermi arc is evident by the presence of multiple gapless spectra (labelled by blue color). Beyond the Fermi arc region, (i) the gapped structure of the EDCs can be realized from $\omega$ locations of their peaks away from $\mu$, and (ii) the peaks of the spectra move progressively away from $\mu$ as the antinode is approached and the peak at the antinode (green curve) is furthest from $\mu$. These observations attest to strong momentum anisotropy of the energy gap in the PG state like in the SC state. Furthermore, the fact that the minimum value of the intensity of each FF-divided gapped EDC is located approximately at $\mu$ is consistent with the PG state possessing PHS.

 The data in Fig. 3a, however, can not be counted as a conclusive proof of PHS in the PG state because the peaks of the gapped spectra for $\omega>0$ can not be readily accessed from ARPES data. This is where the JDOS analysis can be useful. More specifically, as it is previously pointed out, the dispersions of the JDOS peaks are tied to that of the Bogoliubov quasiparticles and hence, the demonstration of $|\mathbf{q}_i (\omega)|=|\mathbf{q}_i (-\omega)|$ in the PG state can be rendered as an independent and direct proof for PHS. It will be ideal to be able to probe such symmetry for all $\mathbf{q}$'s. However, we focus specifically on the JDOS peaks along the Cu-O bond direction because the dispersions of $|\mathbf{q}_1|$  and $|\mathbf{q}_5|$ are clearly visible over a reasonably large energy window in our data. To this end, we concentrate on the line cuts of the constant $\omega$ JDOS patterns from the UD85K sample along the ($\text{q}_\text{x}$, 0) direction as a function of $\omega$ from $-28$ meV (bottommost curve) to $28$ meV (topmost curve) in Fig. 3b.  We also display this plot in terms of a  false-color map in Fig. 3c. As ARPES is predominantly a probe for the occupied states, access to the states at positive values of $\omega$ is available through the thermal tail of the Fermi function. In our data, the ARPES signal for $\omega \ge28$ meV is rather weak for reliable calculations of $\text{C}(\mathbf{q}, \omega)$.  Figs. 3b and 3c highlight approximately symmetric dispersions of $|\mathbf{q}_1|$ and $|\mathbf{q}_5|$ with respect to $\mu$, based on which PHS in the PG state of the UD85K sample can be concluded. Additionally, we show the JDOS intensity maps in Fig. 3e for positive and negative values of $\omega$, whose peak structures seem to mirror each other within experimental error bars as expected in case of the PG phase having PHS.
\begin{figure*}
\includegraphics[width=6.5in]{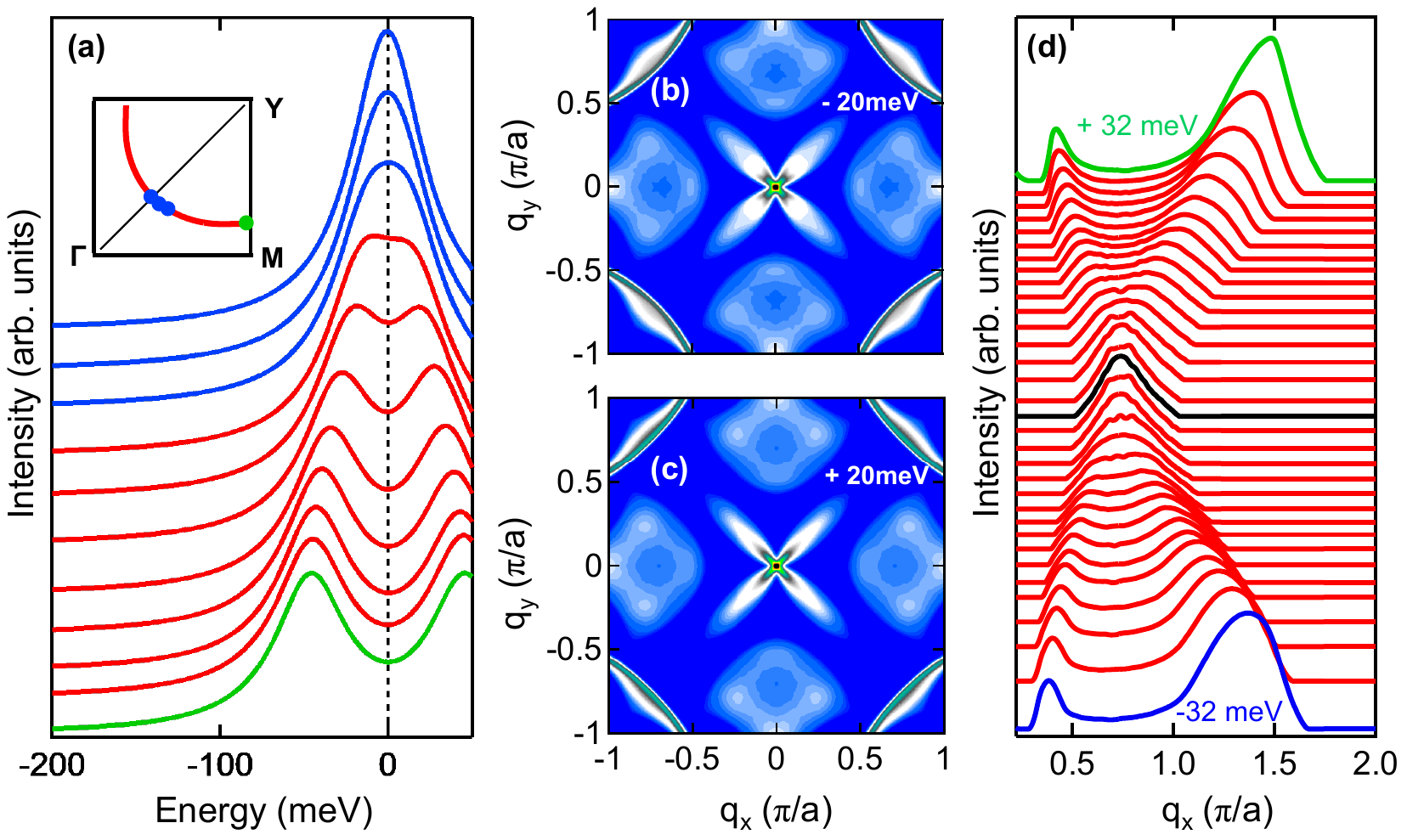}
\caption{\textbf{JDOS patterns and JDOS peak dispersions derived from the broadened $d$-wave SC (BDSC) model.} (a) EDCs, simulated by the BDSC model, along the FS at various momenta between the node (topmost curve) and  the antinode (bottommost curve). Gapless EDCs are marked in blue color. Presence of multiple gapless EDCs along the FS confirms Fermi arc formation in the BDSC model for a finite value of $\Gamma$.  (b, c) Constant $\omega$ JDOS map for (b) $\omega=-20$ meV, and (c) $\omega=+20$ meV. (d) Stacked line cuts  of the JDOS patterns along the Cu-O bond direction as a function of $\omega$ from $-32$ meV (bottommost curve) to $32$ meV (topmost curve). The curves are separated from each other by $2$ meV. $|\mathbf{q}_i (\omega)|=|\mathbf{q}_i (-\omega)|$ for $i=1, 5$ within experimental error bars. JDOS patterns (b, c) and JDOS peak dispersions (d), derived from the BDSC model, agree well with those from the experimental data in the PG phase (Figs, 2 and 3).}
\label{fig:fig4}
\end{figure*}

\subsection{JDOS patterns based on the BDSC model} \label{Sec2D}
 The experimental data from the PG state of our samples in Figs. 1, 2 and 3 provide direct spectroscopic evidences for the presence of (i) PHS and (ii) energy gap with momentum anisotropy similar to that in the SC state. These observations, in totality, point towards a pairing origin of the energy gap, and  thus a phase disordered superconductor picture of the PG state, at least in moderately underdoped \BSCO HTSCs. Previously \cite {R_OCTET_STM3, R_MIKE_ARC_SIM}, we have demonstrated that the key features of the electronic excitations in such a scenario can be effectively captured by using the BDSC model, in which the $\mathbf{k}$ and $\omega$ dependence of the Green's function $\text{G}(\mathbf{k}, \omega)$ can be approximated by the following minimal model: 
\[
\text{G}^{-1}(\mathbf{k}, \omega)=\omega - \epsilon (\mathbf{k}) + i \Gamma - \frac{\big(\Delta(\mathbf{k})\big)^2}{\omega + \epsilon(\mathbf{k}) + i \Gamma}, 
\]
where $\Delta(\mathbf{k})=\dfrac{\Delta_0}{2} \big(\text{cos k}_{\text{x}} - \text{cos k}_{\text{y}}\big)$ is the $d$-wave energy gap with antinodal energy gap $\Delta_0$, $\epsilon(\mathbf{k})$ a tight binding fit to the \BSCO dispersion, and $\Gamma$ a single-particle scattering rate \cite{R_MIKE_ARC_SIM, R_MIKE_CHUBOKOV_ARC_SIM}. This model can qualitatively describe the formation of the so-called Fermi arc for a finite value of $\Gamma$. This can be realized from the plot of the EDCs, derived from the BDSC model, along the FS in Fig. 4a, where the presence of the Fermi arc is highlighted by multiple single-peaked EDCs between the node and antinode along the FS. In Figs. 4b and 4c, we display the JDOS patterns, simulated by the BDSC model, for $\omega=-20$ meV and $\omega=+20$ meV, respectively. Even though this BDSC model is not an exact description of the PG state, the simulated JDOS patterns indeed reproduce the salient features of those obtained from the experimental data, which can be realized from a visual similarity between the JDOS patterns derived from the BDCS model (Figs. 4b, 4c) and experimental data (Fig. 2b). As a specific example, it can be noticed that the characteristic petal-shaped features in the JDOS map from the experimental data are clearly visible in the simulated patterns as well. Moreover, the expected PHS of the JDOS patterns obtained from the BDSC model is evident from the fact that the JDOS peaks at positive and negative energies approximately mirror each other (Figs. 4b, 4c) as like the experimental data (Fig. 3e). To further investigate PHS, we have plotted $\omega$ dependence of  $|\mathbf{q}|$'s along ($\text{q}_{\text{x}}$, 0) for both positive and negative values of $\omega$ in Fig. 4d---the dispersions of both $|\mathbf{q}_1|$ and $|\mathbf{q}_5|$ can be seen to be symmetric with respect to $\mu$ within experimental error bars as like the experimental data (Figs. 3b, 3c).

 In short, the similarity between the peak structures of the JDOS patterns as well as the dispersions of the JDOS peaks from the ARPES data in the PG state and those derived from the BDSC model, provide further support to the picture that the PG state can be effectively expounded as a system of Cooper pairs lacking macroscopic phase coherence to sustain superconductivity. It is worth noting that this is  consistent with previously reported FT-STS analysis of data from underdoped \BSCO \cite{R_DAVIS_PHASE_INCOHERENT_SC}. 
 
 Even though the JDOS patterns based on the BDCS model and the experimental data agree well, it would be insightful to conduct similar comparisons with JDOS patterns derived from various density wave models, which we plan to address in future works.

\section*{Conclusions}
Using JDOS analysis, we have provided direct spectroscopic evidence for PHS in the PG phase of moderately underdoped \BSCO HTSCs, which in turn corroborate the utility of the JDOS analysis to uncover subtle features of electronic excitations that may not be easily detectable from the SDOS analysis of ARPES data. The JDOS patterns and the dispersions of the JDOS peaks together with their PHS in the PG state can be effectively captured by a minimal model involving broadened $d$-wave superconductivity, at least in the moderately underdoped \BSCO HTSCs. 

\section*{Methods}

\subsection*{A. Experimental Details}\label{Sec4A}
The ARPES data used in this paper were collected from single crystal and thin film samples of \BSCO cuprate HTSCs. The $\text{T}_{\text{c}}$ values of the thin film (referred in the text as UD67K) and  the single crystal (referred in the text as UD85K) samples are $67$ K and $85$ K, respectively.
Data were collected using 22 eV photon energy.  Measurements were conducted using Scienta R4000 analyzer at the Synchrotron Radiation Center in Madison, Wisconsin, and at the Swiss Light Source, Paul Scherrer Institute,
Switzerland. The energy resolution was $\sim$15 meV and the momentum resolution was 0.0055 $\si{\angstrom}^{-1}$  along the multiplexing direction of the detector, and 0.02 $\si{\angstrom}^{-1}$ along the perpendicular direction. The data from the single crystal sample were collected in the Y quadrant to minimize superlattice effects. The raw data covered more than one half of the full quadrant of the Brillouin zone for each sample.  We interpolated the data onto a uniform grid and afterwards, reconstructed the whole Brillouin zone by applying reflections appropriate to the symmetry of the CuO$_2$ planes.

\subsection*{B. Data analysis}\label{Sec4B}
Starting from the raw data, we first subtract the approximately $\omega$-independent signal (due to second-order light) at  a sufficiently large positive energy $+|\omega_1|$.
The values of $|\omega_1|$ for UD67K and UD85K samples are $100$ meV and $200$ meV, respectively.
 Afterwards, an “unoccupied” state EDC at a momentum location  far away from Fermi momentum $\text{k}_{\text{F}}$ is taken as an energy-dependent background. The background subtraction is performed by normalizing the background to each EDC at a given negative energy $-|\omega_2|$  and then subtracting it. 
 The values of $|\omega_2|$ for UD67K and UD85K samples are $640$ meV and $570$ meV, respectively. 
The  above-described normalization and background subtraction procedure has been employed in our previous autocorrelation analysis as well in the calculation of dynamic spin susceptibility from the ARPES data \cite{R_UC_SUSCEPTIBILITY}. The resulting intensity maps still show weak superlattice bands, which are not intrinsic to the electronic structure of the CuO$_2$ planes and give rise to features in the JDOS patterns. For this reason, at each energy, intensities below a certain threshold were zeroed to enhance contrast of the constant $\omega$ SDOS intensity maps. 
For the UD67K sample both in the SC and PG states, this threshold was set at 50\%, whereas it was set at a higher value of 55\% for the PG state of the UD85K sample. The above mentioned cutoff values were chosen for the calculations of $\text{C}(\mathbf{q}, \omega)$ in Figs. 1 and 2. 
 As we have established in our previous works \cite {R_OCTET_STM1,R_OCTET_STM2,R_OCTET_STM3} and also found in the current work, the JDOS patterns and the dispersion of the JDOS peaks do not undergo qualitative changes due to these cutoffs. In short, the main effect of thresholding is to enhance the contrast of the structures associated with the main energy bands of the samples.

For the calculations based on the BDSC (Fig.  4), a similar thresholding scheme was necessary. This, however, was not due to the superlattice bands because no such bands were considered in the calculation, but it was necessary to weaken the signal coming from the back-bended  Bogoliubov bands.
In this case, the threshold was set at 30\%.  



\section*{Acknowledgements}
The work was partially supported by the   National Science Foundation under Grant No. DMR-1454304.  The authors have no conflicts to disclose. We acknowledge and thank J. C. Campuzano and Francisco Restrepo for insightful discussions and communications.

\section*{Author contributions}

U.C. designed and conceived the study, U.C. and J.Z. collected the data, N. S., J. Z and U. C analyzed the data, U.C., J. Z and N.S., wrote the manuscript.

\end{document}